\documentclass[fleqn,usenatbib]{rasti}

\usepackage{newtxtext,newtxmath}
\usepackage[T1]{fontenc}
\usepackage{ae,aecompl}
\usepackage{graphicx}
\usepackage{amsmath}
\usepackage{booktabs}
\usepackage{multirow}
\usepackage{xcolor}
\usepackage{hyperref}
\usepackage{orcidlink}

\hypersetup{
	colorlinks=true,
	citecolor=blue,
	urlcolor=blue,
	linkcolor=blue
}

\title[Minimal labels for AGN/SFG]{Stable AGN/SFG classification with minimal labels: K‑means sampling in the MIGHTEE‑COSMOS field}

\author[V. Asadi]{
	Vahid Asadi\,\orcidlink{0009-0005-8897-2385}$^{1}$\thanks{E-mail: vahijd.asadij@gmail.com}
	\\
	$^{1}$Department of Physics, Institute for Advanced Studies in Basic Sciences (IASBS), PO Box 11365-9161, Zanjan, Iran\\
}

\date{Accepted XXX. Received YYY; in original form ZZZ}

\pubyear{2026}

\begin{document}
	\label{firstpage}
	\pagerange{\pageref{firstpage}--\pageref{lastpage}}
	\maketitle
	
	
	\begin{abstract}
		 
	Classifying radio sources as active galactic nuclei (AGN) or star-forming galaxies (SFGs) is essential for deep radio surveys, but obtaining spectroscopic labels for millions of sources is prohibitively expensive. We develop a label-efficient method that classifies radio sources with very few labeled examples while preserving astrophysical population trends. We present a semi-supervised learning (SSL) framework combining repeated K-means clustering with a CatBoost classifier. Using a sample of about 4,500 sources from the MIGHTEE-COSMOS survey, the method applies K-means clustering to unlabeled data in a five-dimensional feature space (the infrared-radio correlation parameter, optical compactness, stellar mass, and two mid-infrared colors), selects the 50 most frequent cluster centroids across repeated runs, and trains a CatBoost classifier on those 50 labeled sources. Evaluated on a held-out test set of 1284 sources, the classifier achieves a weighted F1-score of \(0.9112 \pm 0.0032\) (AGN F1 = \(0.8693 \pm 0.0042\), SFG F1 = \(0.9345 \pm 0.0027\)). The method is stable: over repeated independent runs the weighted F1-score varies with a standard deviation of only 0.0032, compared to \(\sim\)0.018 for random baselines. A fully supervised classifier using all labels achieves a weighted F1-score of 0.9464, within \(\sim\)4\% of this upper-bound performance. Active learning further improves performance to \(0.9284 \pm 0.0051\) at 100 labels. Beyond classification accuracy, the method recovers the known increase in AGN fraction with redshift and stellar mass. Our SSL framework provides a stable, label-efficient method for AGN/SFG classification that preserves astrophysical population trends, making it suitable for large radio surveys.
		
	\end{abstract}
	


\section{Introduction}
Deep radio continuum surveys are powerful probes of galaxy evolution, tracing two fundamental, co-evolving processes: star formation and active galactic nuclei (AGN) accretion. The radio emission from galaxies originates primarily from synchrotron radiation, which is largely unaffected by dust obscuration, providing a uniquely unbiased view of these processes across cosmic time \citep[e.g., ][]{heckman2014coevolution,madau2014cosmic}. Star-forming galaxies (SFGs) dominate the sub-mJy radio source population, with their emission arising from supernova-accelerated cosmic rays. In contrast, at higher flux densities, the radio sky is dominated by AGN-driven emission from relativistic jets and lobes \citep[e.g., ][]{sadler1989low,condon1992radio,miley2008distant}. Distinguishing between these two populations – SFGs and radio AGNs – is therefore a critical prerequisite for exploiting radio surveys to measure cosmic star-formation histories, trace black-hole growth, and understand AGN feedback \citep[e.g.,][]{bonzini2013sub,madau2014cosmic}.

This classification task is growing increasingly urgent with the advent of sensitive, wide-area surveys from precursors to the Square Kilometre Array (SKA), such as the upgraded Karl G. Jansky Very Large Array \citep{perley2011expanded}, the MeerKAT Array \citep{jarvis2017meerkat}, and the Australian Square Kilometre Array Pathfinder \citep{hotan2021australian}. These instruments are detecting millions of radio sources, the majority of which are faint, composite, or ambiguous in nature \citep[e.g., ][]{padovani2016faint}. Traditional multi-wavelength diagnostics (e.g., radio-excess, mid-infrared colors, X-ray luminosity, optical morphology) can provide reliable labels \citep[e.g., ][]{padovani2017two} but are observationally expensive, often incomplete, and subject to redshift-dependent biases \citep[e.g., ][]{whittam2022mightee}.

Consequently, machine learning (ML) has emerged as a vital and transformative tool for scalable astronomical classification. By learning complex patterns from high-dimensional, multi-wavelength datasets, ML algorithms enable the efficient identification and characterization of diverse astronomical populations \citep[e.g., ][]{clarke2020identifying,chaini2023photometric,von2024j,asadi2025leveraging,Asadi_2025,silima2025machine}. However, ML methods rely on high-quality labeled training sets, and obtaining such labels-- especially via spectroscopy--remains costly and time‑consuming.

To address this labeling bottleneck, semi‑supervised learning (SSL) offers a promising alternative by combining a small amount of labeled data with the large pools of unlabeled data that are typical of modern surveys \citep[e.g., ][]{reddy2018semi,van2020survey}. SSL has been successfully applied to various astronomical problems, including AGN classification, supernova identification, redshift estimation, and galaxy morphology \citep[e.g.,][]{lawlor2016mapping,villar2020superraenn,pantoja2022semi,slijepcevic2022radio,ciprijanovic2023deepastrouda,asadi2025semi}.

In this paper, we present a simple clustering‑based SSL framework for AGN/SFG classification. We use a sample of about 4,500 sources from the MIGHTEE‑COSMOS survey \citep{heywood2022mightee}, with five physically motivated features that have been established in the literature (infrared‑radio correlation, optical compactness, stellar mass, and two mid‑infrared colors). The method first applies K‑means clustering \citep{macqueen1967some} to the training set feature space (i.e., using only the features, without any labels) to identify the geometric structure of the data. It then labels only the single source closest to each cluster centroid--the most representative source of each cluster (informed sampling)--and trains a CatBoost classifier \citep{prokhorenkova2018catboost} on those few labeled examples. The resulting classifier is evaluated on a held‑out test set. The goal is to demonstrate that with a small number of carefully selected labels--here 50--one can obtain a stable and reasonably accurate classifier, and that informed sampling significantly reduces variance and minimizes the risk of poor performance. Beyond classification accuracy, we show that the method reproduces known astrophysical population trends, such as the increase in AGN fraction with redshift and stellar mass, with predictions consistent with the true values within the uncertainties.

The paper is organized as follows. Sections~\ref{sec:2} and~\ref{sec:3} describe the MIGHTEE‑COSMOS data and the sample used in this work. Section~\ref{sec:4} presents the SSL framework, and Section~\ref{sec:5} defines the evaluation metrics. Results are given in Section~\ref{sec:6}, including main performance, stability comparison, sensitivity to the number of clusters, a comparison with a fully supervised baseline, and an active learning extension. Section~\ref{sec:7} discusses the findings and their limitations, and Section~\ref{sec:8} concludes.

\section{Data}\label{sec:2}
This work utilizes early science radio continuum data from the MIGHTEE survey--one of the MeerKAT large survey projects--in the COSMOS field \citep{heywood2022mightee}. The COSMOS field was observed for 17.45 hours on source between 2018 and 2019 using MeerKAT's L-band receivers (856–1712, MHz), achieving a thermal noise of 1.7 $\mu$Jy/beam (rising to $\sim$ 4--5 $\mu$Jy/beam at the field center due to classical confusion) and a circular synthesized beam size of 8.6 arcsec $\times$ 8.6 arcsec.

The MIGHTEE-COSMOS multi-wavelength catalogue contains 5223 radio sources with robust host galaxy identifications via K$_\text{s}$-band counterparts from UltraVISTA DR4 \citep{whittam2022mightee,whittam2024mightee}. Optical, near-infrared, and mid-infrared photometry were sourced from HSC-SSP, CFHTLS, UltraVISTA, and Spitzer/IRAC \citep{adams2021evolution}. Far-infrared detections (24--500 $\mu$m) were obtained from Herschel/MIPS and PACS \citep{vaccari2016spitzer,shirley2021help}. Spectroscopic redshifts are available for 2427 sources; photometric redshifts for the remaining 2796 sources were estimated using a hierarchical Bayesian approach \citep{hatfield2022hybrid}. Stellar masses were derived via AGNFITTER spectral energy distribution (SED)-fitting \citep{whittam2022mightee}.

\cite{whittam2022mightee} classified sources as AGNs or SFGs using five diagnostic methods: radio excess, mid-infrared color–color, optical morphology, X-ray luminosity, and very long baseline interferometry brightness temperature. A source was classified as an AGN if it satisfied any one criterion. Sources not meeting AGN criteria were classified as SFGs; however, sources with $\text{z} > 0.5$ undetected in X-rays received the intermediate classification ``probable SFG'' due to redshift-dependent X-ray detection limits.

\section{Data sample}\label{sec:3}
To implement our SSL framework, we define a clean sample based on five physically motivated features: the infrared–radio correlation parameter \(\text{q}_{\mathrm{\text{IR}}}\), the optical compactness \texttt{class\_star}, the stellar mass \(\log(\text{M}/\text{M}_\odot)\), and two mid‑infrared colors: \(\log(\text{S}_{3.6}/\text{S}_{4.5})\) and \(\log(\text{S}_{5.8}/\text{S}_{8.0})\). All quantities are taken directly from the MIGHTEE‑COSMOS catalogue described in Section~\ref{sec:2}.

\begin{figure*}
	\centering
	\includegraphics[width=\linewidth]{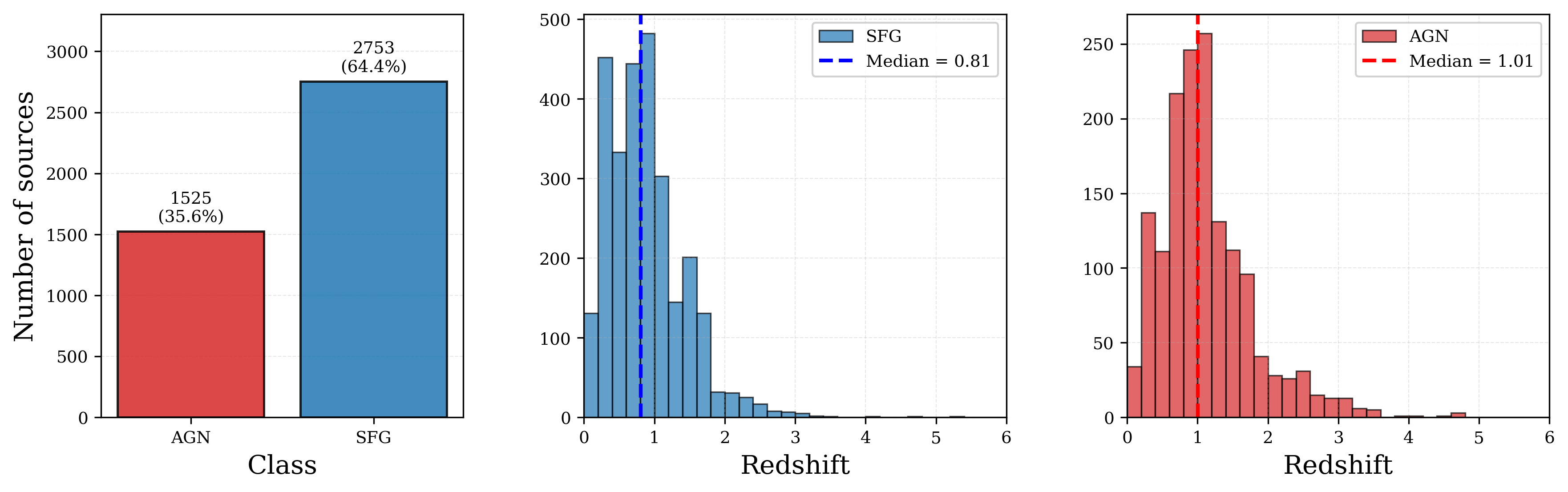}
	\caption{MIGHTEE-COSMOS sample composition and redshift distributions used in this study.}
	\label{fig:sample_composition}
\end{figure*}

We first exclude sources that lack reliable measurements in any of these five features. Flux values flagged as \(-99.9\) (non‑detection) are treated as missing, and any source with at least one missing value is removed. We further require positive detections in the SPLASH bands used to compute the mid‑infrared colors (SPLASH 1 and SPLASH 2 for \(\log(\text{S}_{3.6}/\text{S}_{4.5})\), and SPLASH 3 and SPLASH 4 for \(\log(\text{S}_{5.8}/\text{S}_{8.0})\)); sources with zero or negative flux in any of these bands are removed. Finally, we remove sources originally flagged as `unclass' (``probable SFG'', see Section~\ref{sec:2}) to keep only unambiguous AGN and SFG labels. After these cleaning steps, the final dataset comprises 4278 sources, of which 1525 are AGN and 2753 are star‑forming galaxies. Their redshift distribution and class composition are shown in Figure~\ref{fig:sample_composition}.

The reasons for adopting these five features are rooted in established multi‑wavelength diagnostics. The infrared–radio correlation parameter \(\text{q}_{\mathrm{\text{IR}}}\) captures the tight relation between far-infrared and radio emission in SFGs \citep[e.g., ][]{condon1992radio}; radio excess above this correlation (i.e., low \(\text{q}_{\mathrm{\text{IR}}}\)) is a well-established signature of AGN activity \citep[e.g., ][]{ivison2010blast,delhaize2017vla}. The SExtractor-derived \texttt{class\_star} parameter \citep[e.g., ][]{bertin1996sextractor} probes optical morphology; unobscured AGN with bright unresolved nuclei are more likely to appear as point-like sources, whereas SFGs typically show extended disk-dominated morphologies. Stellar mass \(\log(\text{M}/\text{M}_\odot)\) is included as a contextual prior: AGN preferentially reside in massive galaxies (\(\gtrsim 10^{10} \text{M}_\odot\)), helping to avoid misclassifying low-mass starbursts \citep[e.g., ][]{kauffmann2003host,best2005host}. The two mid‑infrared colors, \(\log(\text{S}_{3.6}/\text{S}_{4.5})\) and \(\log(\text{S}_{5.8}/\text{S}_{8.0})\), trace the hot-dust continuum of the AGN torus and the PAH features in SFGs; together they form the classic color–color wedge diagrams used to select AGN \citep[e.g., ][]{stern2005mid}.

\section{Semi-supervised framework}\label{sec:4}
SSL is a ML approach that exploits both a small labeled subset and a large pool of unlabeled data. This is particularly advantageous when labeled examples are sparse or expensive to obtain. By combining the principles of supervised and unsupervised learning, SSL aims to achieve improved model performance and generalization. Among the many existing approaches \citep[e.g.][]{zhu2005semi,reddy2018semi,van2020survey, asadi2025semi}, we adopt a clustering-based SSL framework. Before describing the full implementation, we briefly review the two core algorithms underlying this approach: K-means clustering \citep{macqueen1967some} and the CatBoost classifier \citep{prokhorenkova2018catboost}.

K-means is an unsupervised clustering algorithm that partitions a dataset into a predefined number of clusters, K. Its objective is to assign data points to clusters such that intra-cluster variance is minimized while inter-cluster separation is maximized. The algorithm begins with randomly initialized centroids and alternates between two steps: (1) assigning each data point to the nearest centroid, and (2) updating each centroid as the mean of the data points assigned to it. These steps are repeated iteratively until convergence, defined by minimal centroid displacement or a maximum iteration limit. Common distance metrics include the Euclidean and Manhattan distances. Despite its efficiency, a key limitation of K-means is its sensitivity to centroid initialization. Random initialization of centroids can lead to convergence to different local minima, resulting in highly variable clustering outcomes across multiple runs \citep[e.g., ][]{celebi2013comparative}.

CatBoost is a gradient-boosting algorithm designed for classification tasks. It constructs an ensemble of decision trees sequentially, where each successive tree is trained to correct the errors of its predecessors, thereby refining the model's predictions iteratively. This process involves incrementally minimizing a loss function, which enhances the model's ability to generalize from the training data. A distinctive feature of CatBoost is its implementation of ordered boosting. This technique directly addresses overfitting--the scenario in which a model memorizes the training data and consequently fails to generalize to unseen examples. By carefully balancing model complexity with the available data, CatBoost's ordered boosting ensures the model remains robust and effective, delivering high performance and stable predictions even on diverse datasets. It achieves this by introducing a more principled and less biased way of calculating gradients, which are the signals the algorithm uses to learn, ultimately resulting in a model that is less prone to memorizing noise in the data.

\begin{figure*}
	\centering
	\includegraphics[width=\linewidth]{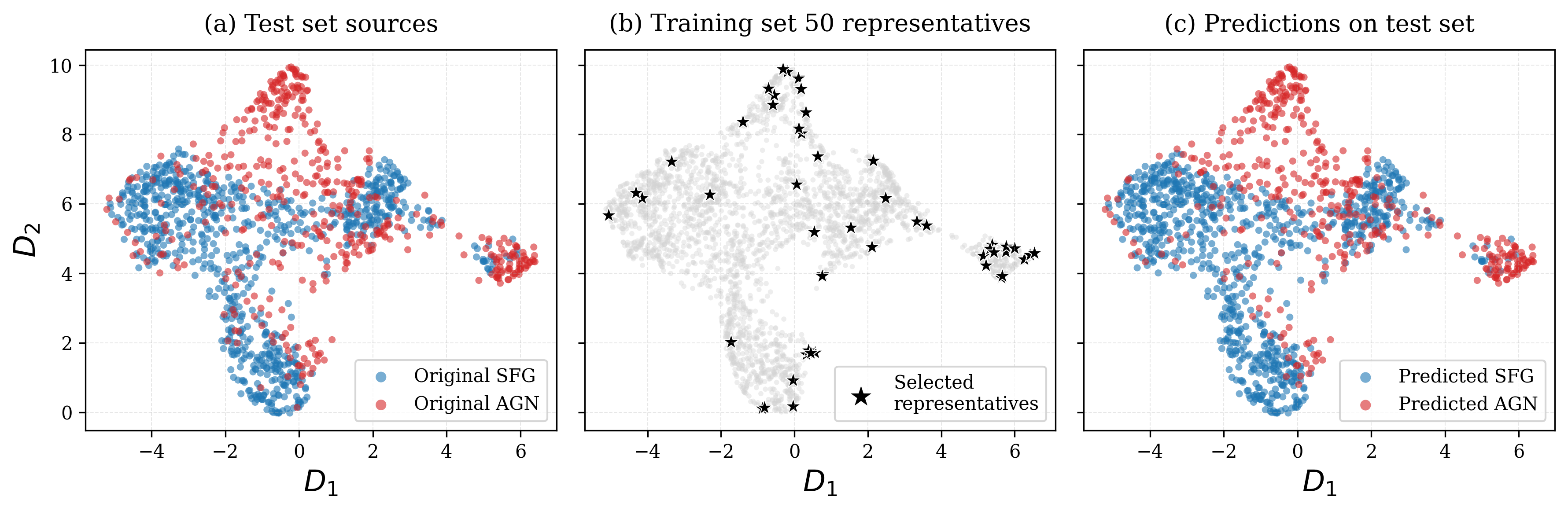}
	\caption{2D UMAP visualization of the SSL workflow. (a) Test set by true class (AGN: red, SFG: blue). (b) Training set with the 50 selected representatives (black stars). (c) Predictions on the test set.}
	\label{fig:umap_train_test_split}
\end{figure*}

\subsection{Informed Selection of Representative Sources}
The core idea of our SSL framework is to select a small set of sources that best represent the geometric structure of the unlabeled dataset. To obtain an unbiased estimate of generalization, we first split the full dataset (4278 sources) into a training set (70\%) and a test set (30\%) using stratified sampling to preserve the AGN/SFG ratio. The training set is then treated as the unlabeled pool for clustering and representative selection; the test set is held out and used only for final evaluation. This procedure ensures that the test sources do not influence the selection of centroids or the training of the classifier.

The training set (2994 sources: 1067 AGN, 1927 SFG) is used for all subsequent steps. The selection of representatives is achieved by repeatedly applying K‑means clustering and retaining the points that most frequently appear as cluster centroids.

\begin{enumerate}
	\item \textit{Feature standardisation}: All five features are standardized to have zero mean and unit variance using the training set statistics. This ensures that each feature contributes equally to distance calculations.
	
	\item \textit{Repeated clustering}: We perform 1000 independent runs of K‑means with K=50 clusters on the standardized training set features. In each run, the source closest to each cluster centroid is recorded.
	
	\item \textit{Selecting the final 50 representatives}: Across the 1000 runs, we count how many times each source in the training set is chosen. The 50 sources with the highest frequencies are retained as the final representative set. This procedure stabilizes the selection against random initialization of K‑means \citep[e.g., ][]{celebi2013comparative} and ensures that the chosen sources are prototypical of the data distribution.
\end{enumerate}

We set the number of clusters to K=50. This value balances labeling effort against coverage of the five‑dimensional feature space. A sensitivity analysis (Section~\ref{sec:6.3}) confirms that performance plateaus around K=50, with only marginal gains for larger values. We therefore fix K=50 for the main analysis.

\subsection{Training and Prediction}
Once the 50 representative sources are identified from the training set, we obtain their ground‑truth labels (simulating an expert labeling campaign, e.g., via spectroscopy or multi‑wavelength diagnostics). A CatBoost classifier is then trained using only these 50 labeled examples. The classifier is configured with its default parameters, as they were found to be sufficient after preliminary hyperparameter optimisation.

After training, the classifier is applied to the held‑out test set (1284 sources: 458 AGN, 826 SFG). This provides an unbiased estimate of how the method would perform on new, unseen data from the same population. In a real‑world application, the same workflow would be used: select 50 sources for follow‑up from the unlabeled catalogue, train the model, and then classify the rest.

To provide an intuitive visualization, we project the feature space onto a two‑dimensional Uniform Manifold Approximation and Projection \citep[UMAP; ][]{mcinnes2018umap} manifold. Figure~\ref{fig:umap_train_test_split} shows this projection. Panel (a) colors the test set by true class (AGN in red, SFG in blue), illustrating the underlying separation. Panel (b) shows the training set with the 50 selected representatives marked as black stars; these are spread across the manifold, capturing the geometric diversity. Panel (c) displays the predictions of the CatBoost on the test set; the color pattern is visually similar to panel (a), indicating that the classifier generalizes well to unseen data.

\begin{figure*}
	\centering
	\includegraphics[width=0.75\linewidth]{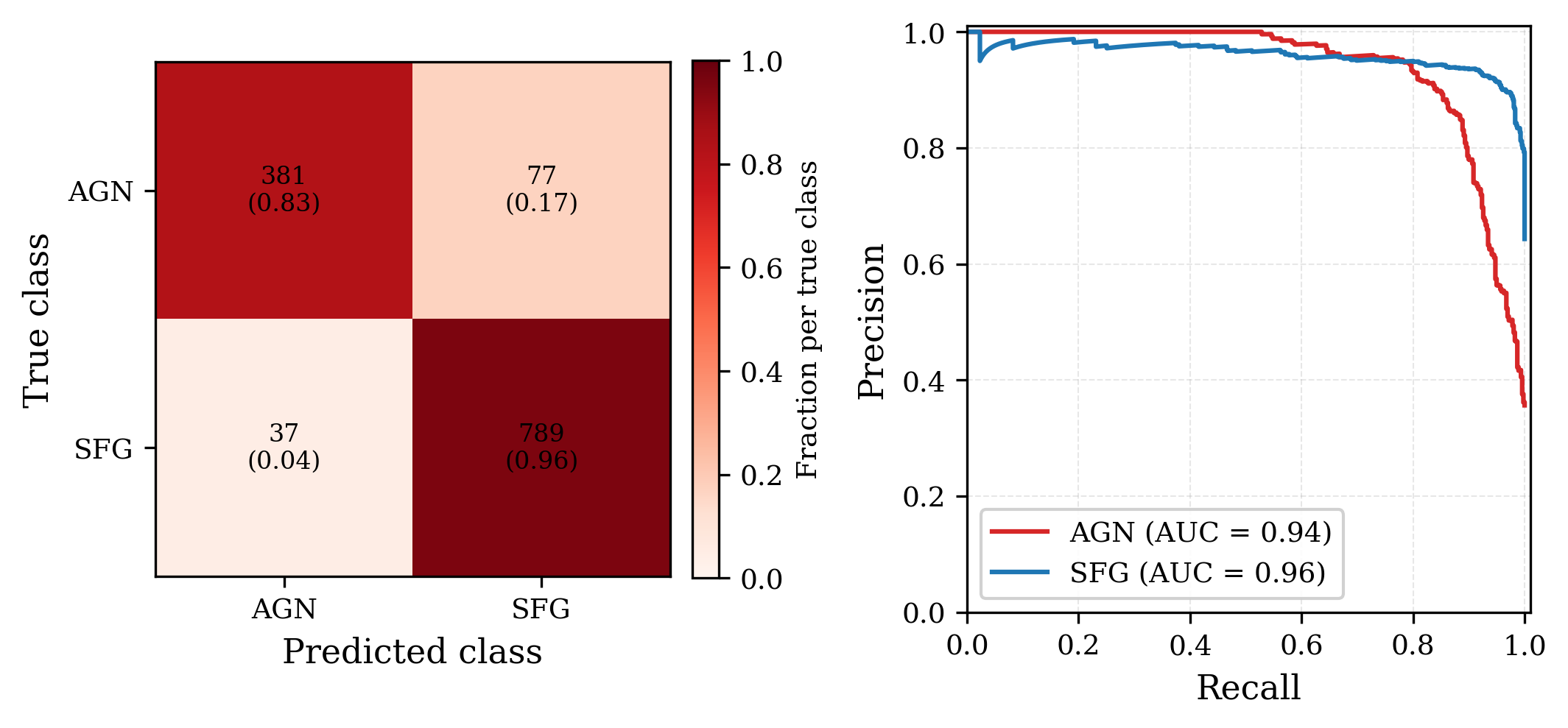}
	\caption{Left panel: Normalized confusion matrix for the SSL classifier at a probability threshold of 0.5, evaluated on the test set (1284 sources). Right panel: Class‑specific precision–recall curves for AGN (red) and SFGs (blue). The area under the curve (AUC) is reported in the legend for each class.}
	\label{fig:confusion_pr_curves}
\end{figure*}

\section{Evaluation Metrics}\label{sec:5}
To evaluate the performance of our classifier on the held‑out test set, we employ standard metrics derived from the confusion matrix \citep{stehman1997selecting}: precision (purity), recall (completeness), and the F1‑score.

The confusion matrix summarizes the performance of a classification model on a set of sources with known labels. It contains four key quantities: True Positives (TP), True Negatives (TN), False Positives (FP), and False Negatives (FN).

Precision measures the accuracy of positive predictions, or the purity of the selected positive class:
\begin{equation}
	\text{Precision} = \frac{\text{TP}}{\text{TP} + \text{FP}}.
	\label{eq:1}
\end{equation}

Recall measures the fraction of all actual positives that are correctly identified, or the completeness of the selection:
\begin{equation}
	\text{Recall} = \frac{\text{TP}}{\text{TP} + \text{FN}}.
	\label{eq:2}
\end{equation}

In the context of this work, a positive prediction corresponds to the identification of an AGN. A high precision therefore indicates a pure sample of AGNs, in which most of the identified candidates are truly AGNs, while a high recall indicates a complete sample, meaning the model successfully recovers a large fraction of the true AGN population.

The F1‑score, defined as the harmonic mean of precision and recall, provides a single metric that balances the trade‑off between purity and completeness:
\begin{equation}
	\text{F1-Score} = 2 \times \frac{\text{Precision} \times \text{Recall}}{\text{Precision} + \text{Recall}}.
	\label{eq:3}
\end{equation}

\section{Results}\label{sec:6}

\subsection{Semi-supervised Performance}\label{sec:6.1}
We evaluate the classification performance of our SSL framework on the held‑out test set (1284 sources: 458 AGN, 826 SFG). The main performance metrics are reported as the mean $\pm$ standard (see Table~\ref{tab:ssl_performance}) deviation over 1000 independent runs, each with different K‑means initializations and CatBoost seeds. Figure~\ref{fig:confusion_pr_curves} shows the normalized confusion matrix and precision–recall curves for a representative run (the one closest to the mean weighted F1-score).

The confusion matrix (left panel) reveals that of the 458 true AGN, the model correctly identifies 381 sources (83\% true positive rate), misclassifying 77 as SFGs. Of the 826 true SFGs, 789 are correctly classified (96\% true positive rate), with 37 false positives. Inspection of the confusion matrix shows that AGN are more often misclassified as SFGs (17\%) than SFGs as AGN (4\%). This is consistent with the known difficulty of identifying low‑luminosity or obscured AGN, which can overlap substantially with SFGs in radio and multi-wavelength feature space.

To gain insight into the nature of misclassified sources, we compared their median physical properties to those of correctly classified sources (Table~\ref{tab:misclassification}). Misclassified AGN tend to be at lower redshift, have lower stellar mass, and exhibit weaker radio excess (higher \(\text{q}_{\mathrm{\text{IR}}}\)) than correctly classified AGN. This suggests that they are predominantly low‑luminosity, radio‑quiet AGN whose signatures are less distinct, causing them to be confused with SFGs. Conversely, misclassified SFGs tend to be at higher redshift, more massive, and have stronger radio emission (lower \(\text{q}_{\mathrm{\text{IR}}}\)) than correctly classified SFGs. They also show slightly higher optical compactness. These are likely composite sources or high‑redshift starbursts whose radio emission overlaps with the AGN locus. The optical morphology parameter \texttt{class\_star} does not strongly separate misclassified from correctly classified sources, indicating that the confusion arises primarily from the radio–IR diagnostic space.

\begin{table}
	\centering
	\caption{Median properties of correctly and misclassified sources.}
	\label{tab:misclassification}                    
	\begin{tabular}{c|cccc}
		\hline\hline
		Group & $\text{z}$ & $\log(\text{M}_\star/\text{M}_\odot)$ & $\text{q}_{\rm \text{IR}}$ & \texttt{class\_star} \\
		\hline
		AGN (miscl.) & 0.838 & 10.488 & 2.312 & 0.028 \\
		AGN (corr.)  & 1.025 & 10.627 & 1.635 & 0.029 \\
		SFG (miscl.) & 1.368 & 10.782 & 1.978 & 0.029 \\
		SFG (corr.)  & 0.758 & 10.423 & 2.374 & 0.006 \\
		\hline
	\end{tabular}
\end{table}

The precision–recall curves (right panel) give area under the curve (AUC) values of 0.94 for AGN and 0.96 for SFG, indicating that the classifier maintains good discrimination across a range of thresholds.

The corresponding per‑class metrics, averaged over 1000 independent runs, are listed in Table~\ref{tab:ssl_performance}. The SSL classifier achieves a weighted F1‑score of \(0.9112 \pm 0.0032\) (mean $\pm$ std), with AGN F1-score = \(0.8693 \pm 0.0042\) and SFG F1-score = \(0.9345 \pm 0.0027\). These results indicate that a classifier trained on only 50 strategically selected sources generalizes well to unseen data from the same population, with high stability across different initializations.

\begin{table}
	\centering
	\caption{Classification performance of the SSL classifier on the test set (mean $\pm$ std over 1000 runs).}
	\setlength{\tabcolsep}{5pt}
	\label{tab:ssl_performance}
	\begin{tabular}{l|ccc}
		\hline\hline
		Class & Precision & Recall & F1-score \\
		\hline
		AGN   & \(0.9331 \pm 0.0139\) & \(0.8139 \pm 0.0079\) & \(0.8693 \pm 0.0042\) \\
		SFG   & \(0.9037 \pm 0.0033\) & \(0.9675 \pm 0.0074\) & \(0.9345 \pm 0.0027\) \\
		\hline
		Weighted F1& \multicolumn{2}{c}{} & \(0.9112 \pm 0.0032\) \\
		\hline
	\end{tabular}
\end{table}

\subsection{Stability Comparison: Informed versus random sampling}\label{sec:6.2}
To assess the reliability of our SSL framework, we compare it against a random baseline that follows the same procedure except for the selection of representatives. Instead of K‑means centroids, we randomly select 50 distinct sources without replacement per outer run. Both methods then train a CatBoost classifier on those 50 labeled sources and evaluate on the held‑out test set. We repeat the entire selection–training–evaluation process 1000 times (outer loops) for both methods to measure stability.

\begin{figure}
	\centering
	\includegraphics[width=\linewidth]{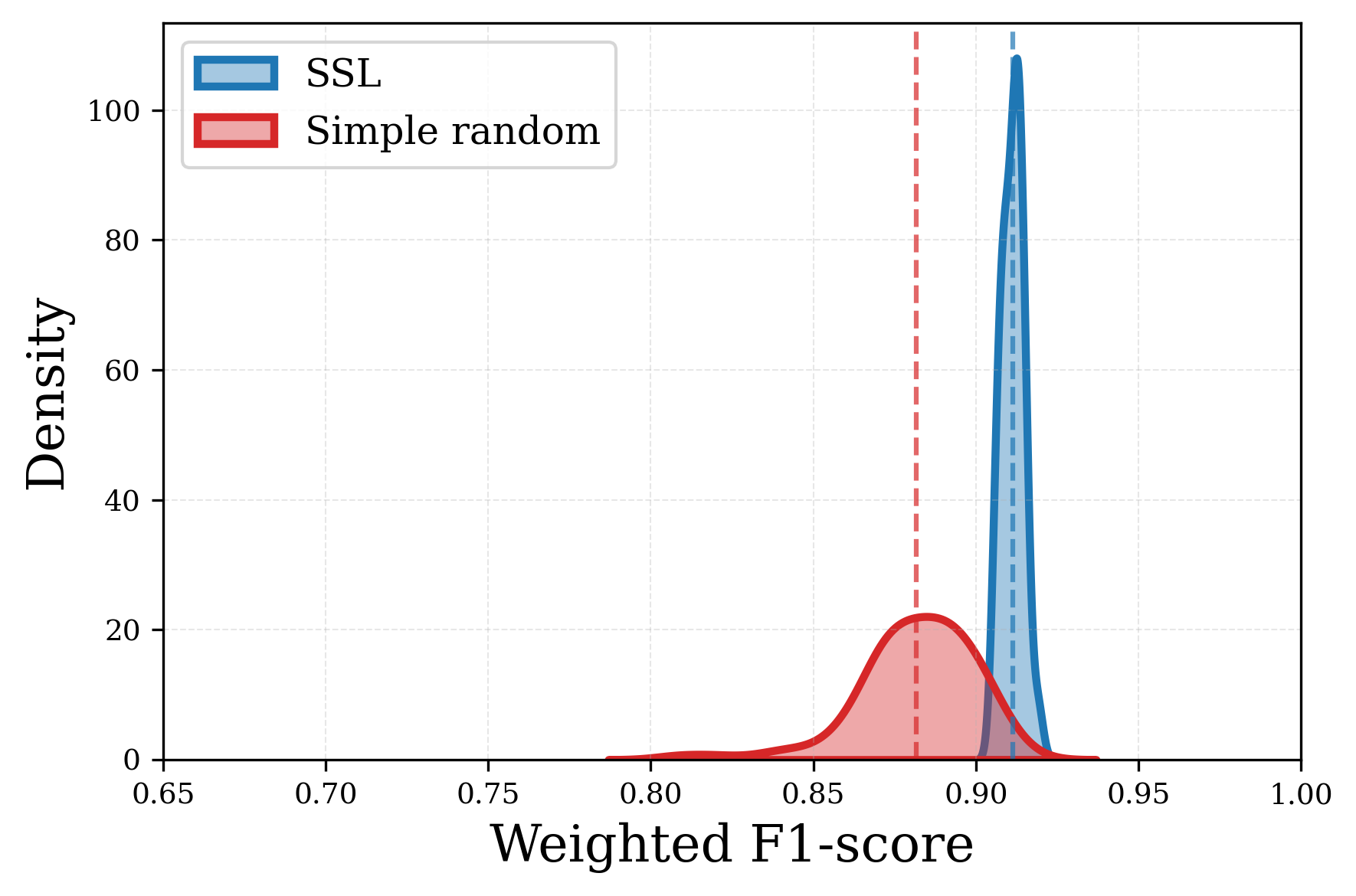}
	\caption{Kernel density estimates of the weighted F1‑score distribution over 1000 outer runs for SSL informed (blue) and simple random (red). The SSL method produces a narrow peak centred at 0.91, while the random baseline shows a broad distribution with a long left tail down to $\sim$0.81.}
	\label{fig:stability_kde}
\end{figure}

Figure~\ref{fig:stability_kde} shows kernel density estimates (KDE) of the weighted F1‑score distributions for the two methods. The SSL method produces a narrow, sharply peaked distribution centered near 0.91, with negligible probability of falling below 0.90. In contrast, the random baseline exhibits a broad distribution with a pronounced left tail that extends down to $\sim$0.81, indicating a real risk of poor performance. This visual difference underscores the reliability of the SSL approach.

Table~\ref{tab:stability_comparison} provides the numerical summary. The SSL method achieves a weighted F1‑score of \(0.9112 \pm 0.0032\), a minimum of \(0.9060\), and a 95\% confidence interval \([0.9060, 0.9166]\). The simple random baseline yields \(0.8818 \pm 0.0181\), a minimum as low as \(0.8086\), and a 95\% interval \([0.8392, 0.9083]\). The standard deviation of the SSL method is nearly six times smaller than that of the random baseline, and its worst‑case performance is substantially higher (0.9060 vs 0.8086).

Per‑class stability follows the same pattern (see Table~\ref{tab:stability_comparison}). For AGN, SSL gives an F1‑score of \(0.8693 \pm 0.0042\) versus \(0.8208 \pm 0.0294\) for random. For SFG, SSL yields \(0.9345 \pm 0.0027\) while random gives \(0.9156 \pm 0.0129\). The recall for AGN is particularly unstable under random selection (\(0.7393 \pm 0.0543\)) compared to SSL (\(0.8139 \pm 0.0079\)).

\begin{table}
	\centering
	\caption{Stability comparison between SSL informed sampling and simple random sampling (50 labels).}
	\label{tab:stability_comparison}
	\begin{tabular}{l|cc}
		\hline
		\hline
		Metric & SSL & Simple Random \\
		\hline
		Weighted F1 (mean$\pm$std) & \(0.9112 \pm 0.0032\) & \(0.8818 \pm 0.0181\) \\
		Weighted F1 (min, max) & \(0.9060,\;0.9189\) & \(0.8086,\;0.9153\) \\
		Weighted F1 (95\% CI) & \([0.9060, 0.9166]\) & \([0.8392, 0.9083]\) \\
		\hline
		AGN F1 & \(0.8693 \pm 0.0042\) & \(0.8208 \pm 0.0294\) \\
		AGN precision & \(0.9331 \pm 0.0139\) & \(0.9292 \pm 0.0507\) \\
		AGN recall & \(0.8139 \pm 0.0079\) & \(0.7393 \pm 0.0543\) \\
		\hline
		SFG F1 & \(0.9345 \pm 0.0027\) & \(0.9156 \pm 0.0129\) \\
		SFG precision & \(0.9037 \pm 0.0033\) & \(0.8707 \pm 0.0216\) \\
		SFG recall & \(0.9675 \pm 0.0074\) & \(0.9663 \pm 0.0280\) \\
		\hline
	\end{tabular}
\end{table}

These results demonstrate that the SSL method not only achieves a higher average performance but also substantially reduces the risk of poor performance compared to random sampling. The low variance and narrow confidence intervals make the method suitable for practical surveys where a single reliable set of labels must be chosen.

\subsection{Sensitivity to the Number of Clusters}\label{sec:6.3}

\begin{figure}
	\centering
	\includegraphics[width=\linewidth]{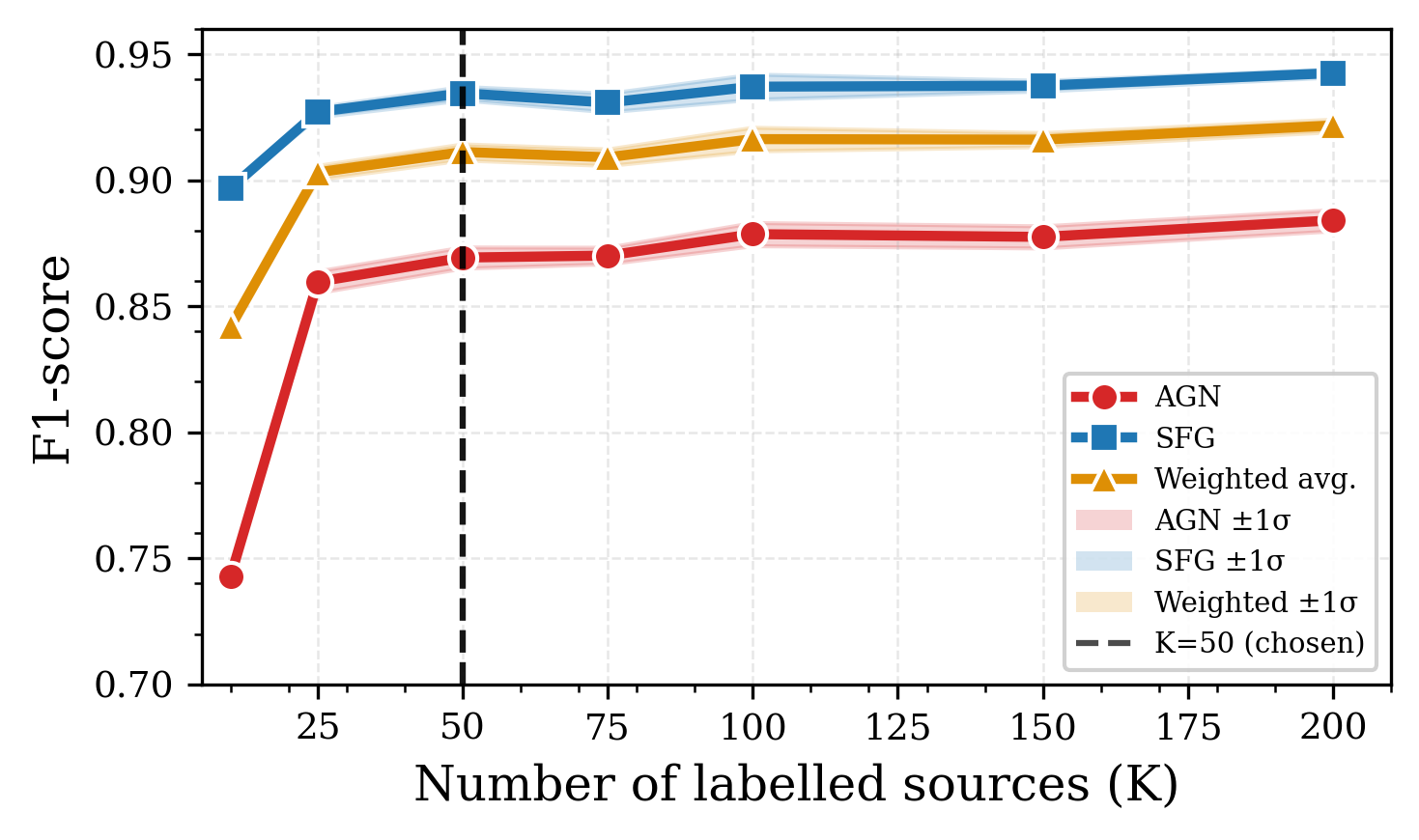}
	\caption{F1‑scores as a function of the number of clusters K (i.e., the number of labeled sources). AGN (red circles), SFG (blue squares), and weighted F1-score (orange triangles) are shown with shaded regions indicating \(\pm1\sigma\) from 1000 independent runs. The dashed vertical line marks the chosen value K=50.}
	\label{fig:k_sensitivity}
\end{figure}

The number of clusters K determines how many sources are selected for labeling. To test the sensitivity of our SSL framework to this parameter, we ran the pipeline with K ranging from 10 to 200, using the same training/test split and evaluating on the held‑out test set. Figure~\ref{fig:k_sensitivity} shows the resulting F1‑scores for AGN, SFG, and the weighted average, with shaded regions representing the standard deviation over the independent runs.

Performance improves rapidly from K=10 (weighted F1-score = 0.842) to K=25 (weighted F1-score = 0.903). From K=25 to K=100, the gain is modest: \(0.903 \to 0.916\). Beyond K=100, the improvement is marginal, with K=200 reaching 0.922. We choose K=50 as a practical trade‑off between labeling effort and classification performance, consistent with the main analysis. The standard deviation is consistently small (\(\sigma < 0.005\)) across all \(\text{K} \ge 25\).

It is worth noting that with only 25 labels the classifier already reaches a weighted F1-score of 0.903, comparable to the 0.911 with 50 labels. This indicates that the method remains effective even at very low label budgets.

\subsection{Comparison with a Fully Supervised Baseline}\label{sec:6.4}

To contextualise the performance of our SSL framework, we trained a fully supervised CatBoost classifier using all 2994 labeled sources in the training set and evaluated it on the same held‑out test set. This provides an upper bound on the performance achievable with this feature set and classifier. To quantify the stability of the supervised model, we repeated the training 1000 times with different random seeds.

Table~\ref{tab:supervised_upper_bound} summarizes the results. The fully supervised model achieves a weighted F1‑score of \(0.9464 \pm 0.0000\), with AGN F1-score = \(0.9220 \pm 0.0000\) and SFG F1-score = \(0.9599 \pm 0.0000\). Our SSL method with only 50 labels reaches a weighted F1‑score of \(0.9112 \pm 0.0032\) (Table~\ref{tab:ssl_performance}), which is within \(\sim\)4\% of the supervised upper bound. This demonstrates that the 50 representatives selected via informed sampling capture a substantial fraction of the discriminative information available in the feature space, achieving performance competitive with the fully supervised model while using only a tiny fraction of the labels (\(<2\%\) of the training set).

\begin{table}
	\centering
	\caption{Performance of the fully supervised CatBoost classifier trained on all 2994 training sources.}
	\setlength{\tabcolsep}{5pt}
	\label{tab:supervised_upper_bound}
	\begin{tabular}{l|ccc}
		\hline\hline
		Class & Precision & Recall & F1-score \\
		\hline
		AGN   & \(0.971 \pm 0.000\) & \(0.878 \pm 0.000\) & \(0.922 \pm 0.000\) \\
		SFG   & \(0.936 \pm 0.000\) & \(0.986 \pm 0.000\) & \(0.960 \pm 0.000\) \\
		\hline
		Weighted F1 & \multicolumn{2}{c}{} & \(0.9464 \pm 0.000\) \\
		\hline
	\end{tabular}
\end{table}

\subsection{Active Learning Extension}\label{sec:6.5}
We investigated whether the performance of our SSL method could be improved further by incorporating active learning, a strategy where the classifier iteratively selects the most informative unlabeled sources for labeling, thereby maximizing performance gains per new label \citep[e.g.,][]{settles2009active}. Starting from the 50 representatives selected via repeated K‑means centroids, we iteratively queried the 10 most uncertain sources in the unlabeled pool (using CatBoost prediction probabilities), added their labels, and retrained the classifier. This procedure was repeated until a total of 100 labels was reached. To assess stability, the entire active learning pipeline was repeated 1000 times.

\begin{figure}
	\centering
	\includegraphics[width=\linewidth]{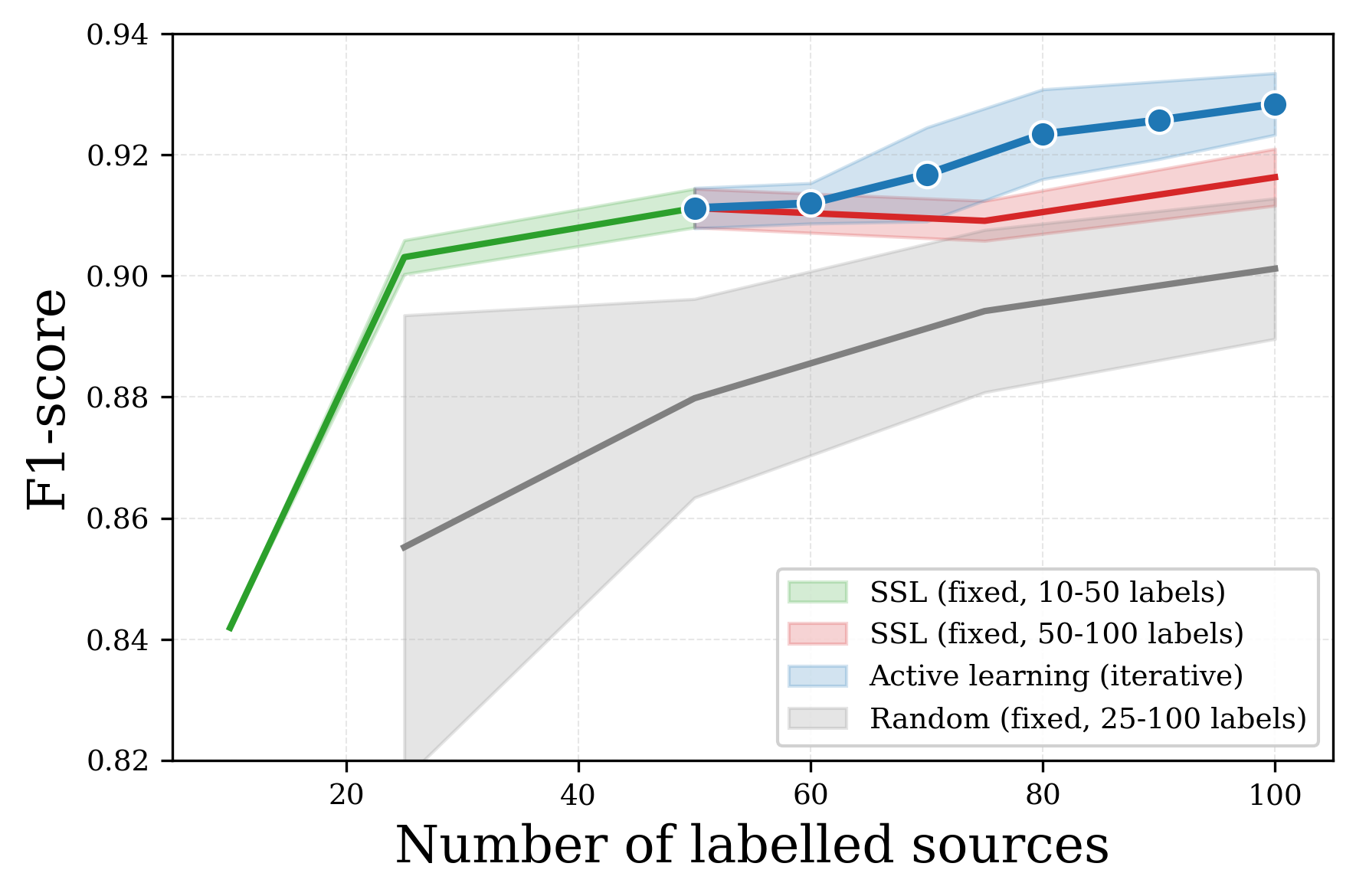}
	\caption{Weighted F1-score as a function of the number of labeled sources for active learning (blue), fixed SSL (green and red), and random sampling (gray). Shaded regions represent \(\pm1\sigma\). Active learning starts from 50 labels and consistently outperforms the fixed SSL method at the same label budget.}
	\label{fig:active_learning_curve}
\end{figure}

Figure~\ref{fig:active_learning_curve} shows the learning curve of the active learning process, with shaded regions representing \(\pm1\sigma\) over 1000 independent runs. The performance increases steadily from 50 to 100 labels. For comparison, we also show the performance of the fixed SSL method at 10, 25, 50, 75, and 100 labels (shown in two segments: 10--50 and 50--100), as well as the random sampling baseline at 25, 50, 75, and 100 labels.

\begin{table*}
	\centering
	\caption{Comparison of active learning, fixed SSL, and random sampling at 100 labels.}
	\label{tab:active_learning_comparison}
	\begin{tabular}{lccc}
		\hline\hline
		Method & Weighted F1-score & AGN F1-score & SFG F1-score \\
		\hline
		Random sampling & \(0.8987 \pm 0.0119\) & \(0.8483 \pm 0.0197\) & \(0.9266 \pm 0.0081\) \\
		SSL fixed & \(0.9163 \pm 0.0047\) & \(0.8786 \pm 0.0046\) & \(0.9371 \pm 0.0050\) \\
		SSL + Active learning & \(\mathbf{0.9284 \pm 0.0051}\) & \(\mathbf{0.8948 \pm 0.0080}\) & \(\mathbf{0.9470 \pm 0.0036}\) \\
		\hline
	\end{tabular}
\end{table*}

\begin{figure*}
	\centering
	\includegraphics[width=\linewidth]{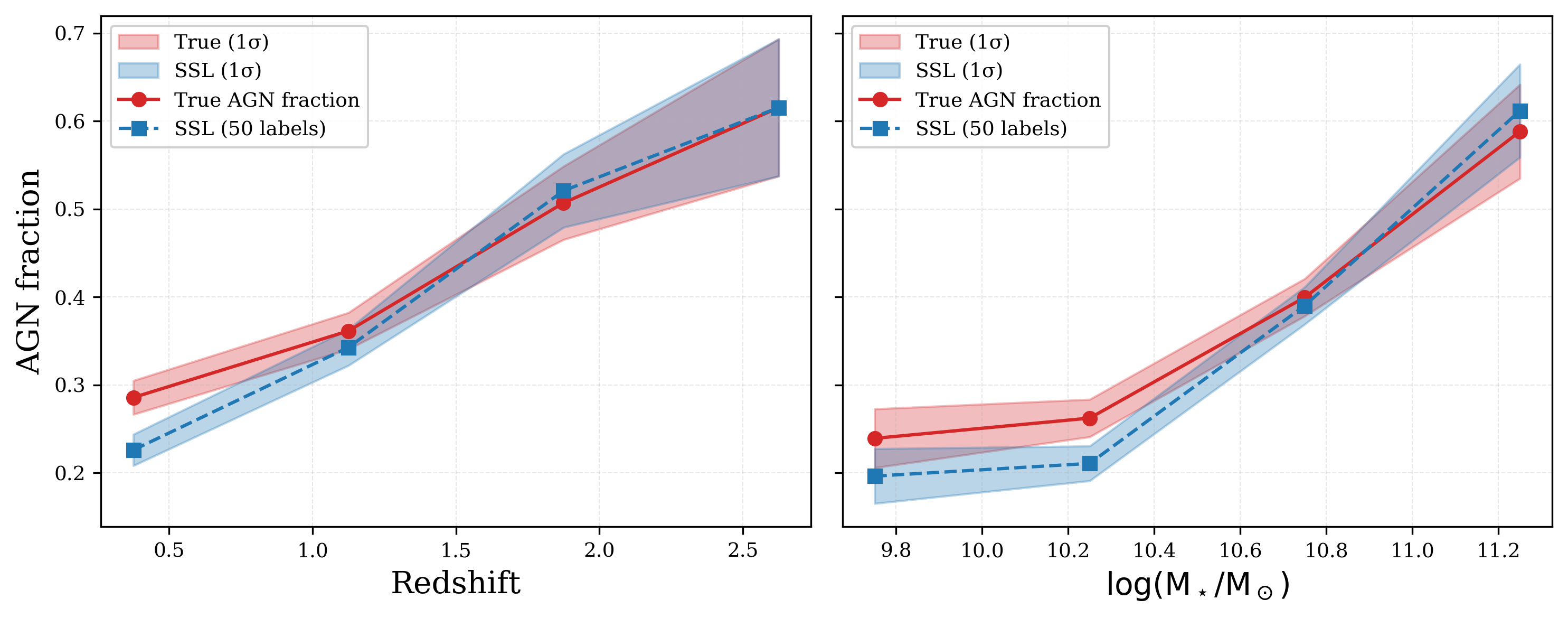}
	\caption{Population-level recovery of the SSL method. Left panel: True (red) and SSL-predicted (blue) AGN fraction as a function of redshift. Right panel: True (red) and SSL-predicted (blue) AGN fraction as a function of stellar mass. Shaded regions represent \(1\sigma\) binomial uncertainties.}
	\label{fig:agn_fraction_population}
\end{figure*}

Table~\ref{tab:active_learning_comparison} summarizes the results at 100 labels. Active learning achieves a weighted F1‑score of \(0.9284 \pm 0.0051\), compared to \(0.9163 \pm 0.0047\) for the fixed SSL method and \(0.8987 \pm 0.0119\) for random sampling. The improvement over fixed SSL is \(0.0121\) in weighted F1-score, with comparable standard deviations (\(0.0051\) vs \(0.0047\)), indicating that the improvement is stable. Active learning also significantly outperforms random sampling, with a gain of \(0.0297\) in weighted F1-score, and the minimum performance of active learning (\(0.9138\)) is substantially higher than that of random sampling (\(0.8687\)). The per‑class performance follows the same pattern: active learning achieves AGN F1-score = \(0.8948 \pm 0.0080\) and SFG F1-score = \(0.9470 \pm 0.0036\), compared to \(0.8786 \pm 0.0046\) and \(0.9371 \pm 0.0050\) for fixed SSL.

These results demonstrate that active learning can effectively improve the performance of the SSL method without requiring additional labels beyond the fixed budget. The iterative querying of uncertain sources allows the classifier to focus on the most informative examples, leading to better performance with the same number of labels.

\section{Discussion}\label{sec:7}
Our results demonstrate that a classifier trained on only 50 strategically selected sources can distinguish AGN from SFGs with reasonable accuracy (weighted F1-score = 0.911) and high stability ($\sigma$ = 0.0032). This confirms that informed sampling via repeated K‑means centroids is an effective strategy for reducing labelling effort in radio surveys, particularly when a single, reliable model is needed. The remainder of this section interprets these findings, explains the geometric basis of the success of informed sampling, presents an astrophysical validation based on the recovery of the AGN fraction with redshift, discusses the practical application of the method to new surveys, and outlines limitations and future directions.

\subsection{Why Informed Sampling Works}

The success of our method can be understood by examining the geometry of the feature space. The five physical diagnostics used here -- \(\text{q}_{\mathrm{\text{IR}}}\), \texttt{class\_star}, \(\log(\text{M}/\text{M}_\odot)\), and the two mid‑infrared colors \(\log(\text{S}_{3.6}/\text{S}_{4.5})\) and \(\log(\text{S}_{5.8}/\text{S}_{8.0})\) -- produce a low‑dimensional manifold in which AGN and SFGs occupy largely distinct regions (as illustrated by the UMAP projection in Figure~\ref{fig:umap_train_test_split}). By applying K‑means repeatedly and retaining the sources that most frequently appear as cluster centroids, we effectively identify natural representatives of the main clusters. These representatives are chosen without any label information; they are simply the points that lie closest to the cluster centers in the unlabeled feature space. Once these 50 representative sources are labeled (simulating a realistic follow‑up campaign), a CatBoost classifier trained on them can propagate the labels to the remaining sources. Because the representatives are selected to cover the main clusters of the feature space, the classifier generalizes well to new sources from the same distribution on the MIGHTEE‑COSMOS dataset, as demonstrated by the held‑out test set performance. This geometric guidance is the key difference from random sampling, which has no preference for cluster centres and therefore can miss critical regions, leading to higher variance and, as shown in Section~\ref{sec:6.2}, a real risk of poor performance.

\subsection{Recovery of Population-level Trends}
An important test of any classification method is whether it preserves the known physical relationships in the data. To evaluate this, we compared the predicted AGN fraction from our 50‑label SSL model to the true AGN fraction as a function of redshift and stellar mass (Figure~\ref{fig:agn_fraction_population}).

Figure~\ref{fig:agn_fraction_population} (left panel) shows the AGN fraction as a function of redshift. The model captures the overall increase in AGN fraction with redshift, which is a well‑known evolutionary trend. The slight under-prediction at low redshift (\(\text{z} \sim 0.4\)) is consistent with the misclassification analysis in Section~\ref{sec:6.1}, which showed that misclassified AGN tend to be at lower redshift and have weaker radio signatures.

Figure~\ref{fig:agn_fraction_population} (right panel) shows the AGN fraction as a function of stellar mass. The SSL model reproduces the known increase in AGN fraction with stellar mass, with good agreement across the mass range. At intermediate mass (\(\log(\text{M}_\star/\mathrm{M}_\odot) \sim 10.7\)), the predicted and true fractions converge almost perfectly at $\sim$0.40. At low masses (\(\log(\text{M}_\star/\mathrm{M}_\odot) \sim 9.8\)--10.2), the model slightly under-predicts the AGN fraction, while at high masses (\(\log(\text{M}_\star/\mathrm{M}_\odot) \sim 11.2\)), it slightly over-predicts. Critically, the \(1\sigma\) uncertainty bands overlap across almost the entire mass range, indicating that the predicted values are within the statistical uncertainty of the true values.

Together, these results demonstrate that the SSL method not only achieves strong source‑level performance but also reproduces the expected population‑level trends in AGN incidence across redshift and stellar mass.

\subsection{Application to New Surveys}
In a practical deployment scenario, the SSL pipeline would be applied directly to an unlabeled catalogue: K‑means clustering would be run on the full feature space of the survey, 50 representative sources would be selected for follow‑up labeling, and the trained classifier would then be used to predict labels for the remaining sources. In this work, we used a train/test split (70/30) to obtain an unbiased estimate of generalization performance, but the method itself is designed for the single‑pass workflow typical of large surveys.

If a new survey can provide the same five physical features (\(\text{q}_{\mathrm{\text{IR}}}\), \texttt{class\_star}, \(\log(\text{M}/\text{M}_\odot)\), \(\log(\text{S}_{3.6}/\text{S}_{4.5})\), and \(\log(\text{S}_{5.8}/\text{S}_{8.0})\)) with similar measurement quality (e.g., comparable depth, resolution, and flux calibration), the classifier trained on our original 50 labels could, in principle, be applied directly, provided the feature distributions are similar. However, if the new survey differs substantially (e.g., different selection function, significantly different redshift range affecting the interpretation of \(\text{q}_{\mathrm{\text{IR}}}\), or missing some required bands), it is better to re‑run the SSL pipeline on the new data. This requires only a minimal number of new labels and the same five features (if available).

\subsection{Limitations and Future Directions}

Our SSL framework, like any classifier, depends on the separability of the target classes in the feature space. However, because it uses only 50 labeled sources, it is more sensitive to feature quality than a fully supervised method that might overcome moderate noise with abundant labels. If the input features do not clearly separate AGN from SFGs – for example, if one uses raw photometric colors that mix the two populations – the geometric structure becomes ambiguous, the K‑means centroids may fall in mixed regions, and the resulting classifier will suffer. In our work, we deliberately chose five well‑established physical diagnostics that are known to provide good separation.

The five features used here require multi‑wavelength data: far‑infrared (Herschel) and radio data for \(\text{q}_{\mathrm{\text{IR}}}\), mid‑infrared (Spitzer/IRAC) for the two MIR colors, deep optical imaging for \texttt{class\_star}, and multi‑band photometry for SED-fitting to obtain \(\log(\text{M}/\text{M}_\odot)\). For surveys lacking such ancillary data, the method cannot be directly applied. However, for deep extra-galactic fields with rich multi‑wavelength coverage (e.g., COSMOS, XMM‑LSS), these features are available.

We also note that the stability analysis in Section~\ref{sec:6.2} focuses on the variability arising from the stochastic components of the SSL algorithm---namely, K‑means initialization and CatBoost randomness---and is conditional on a fixed train/test split. This is a deliberate choice, as our goal is to quantify the algorithmic stability of the informed sampling procedure rather than the uncertainty due to alternative data partitions. A more comprehensive assessment of generalization uncertainty would require repeated splits or cross‑validation.

For future work, the same SSL framework could be applied to other classification problems in astronomy where labels are scarce and a discriminative feature space can be defined. Moreover, a more thorough validation on independent real survey fields would strengthen the generalizability claims. 

\section{Conclusions}\label{sec:8}
We have presented a clustering‑based SSL framework for classifying radio sources as AGN or SFGs using few labeled examples. The method selects 50 representative sources via repeated K‑means clustering (taking the most frequent centroids across 1000 runs) and trains a CatBoost classifier on those 50 labels. The main results can be summarized as follows:

\begin{itemize}
	\item[\textbullet] With only 50 carefully selected labels, the SSL classifier achieves a weighted F1‑score of \(0.9112 \pm 0.0032\) on a held‑out test set of 1284 sources from the MIGHTEE‑COSMOS field. The per‑class performance is reasonable: AGN reach an F1‑score of \(0.8693 \pm 0.0042\), while SFG reach an F1‑score of \(0.9345 \pm 0.0027\).
	
	\item[\textbullet] The SSL strategy is highly stable: over 1000 independent repetitions, the weighted F1‑score varies with a standard deviation of only 0.0032, and the worst performance observed is 0.906. In contrast, random sampling with 50 labels gives a mean F1‑score of \(0.8818 \pm 0.0181\) and a minimum as low as 0.8086, indicating a real risk of poor performance.
	
	\item[\textbullet] A fully supervised CatBoost classifier trained on all 2994 training labels achieves a weighted F1‑score of 0.9464 on the same test set. Our SSL method reaches 0.9112 with 50 labels, which is within \(\sim\)4\% of this upper-bound performance, indicating that informed sampling captures a substantial fraction of the discriminative information with a small fraction of the labels (\(<2\%\) of the training set).
	
	\item[\textbullet] The integration of active learning further improves performance: starting from 50 SSL labels and iteratively querying 10 uncertain sources at a time, the classifier reaches a weighted F1‑score of \(0.9284 \pm 0.0051\) at 100 labels, an improvement of 0.0121 over the fixed SSL method at the same label budget.
	
	\item[\textbullet] For a new survey that can measure the same five physical features (\(\text{q}_{\mathrm{\text{IR}}}\), \texttt{class\_star}, \(\log(\text{M}/\text{M}_\odot)\), \(\log(\text{S}_{3.6}/\text{S}_{4.5})\), and \(\log(\text{S}_{5.8}/\text{S}_{8.0})\)) with similar quality, the existing model could be applied directly, provided the feature distributions are similar. If the population differs substantially, re‑running the pipeline on the new data requires only a minimal number of new labels.
\end{itemize}

\bibliography{ref}

\end{document}